\documentclass[twocolumn,twoside,preprintnumbers,amsmath,amssymb,pacs]{revtex4}
\usepackage{epsfig}
\usepackage{graphicx}
\usepackage{dcolumn}
\usepackage{bm}

\usepackage{fancyhdr}

\usepackage{pslatex}

\begin{document}

\title{{\Large Lattice Results for the QCD Phase Transition}}

\author{Tereza Mendes}

\affiliation{Instituto de F\'{\i}sica de S\~ao Carlos,
  Universidade de S\~ao Paulo, \\
  Caixa Postal 369, 13560-970, S\~ao Carlos, SP, Brazil  }


\received{on 24 March, 2006}

\begin{abstract}

We discuss recent results for the phase transition in finite-temperature QCD
from numerical (Monte Carlo) simulations of the lattice-regularized theory.
Emphasis is given to the case of two degenerate light-quark flavors. The order
of the transition in this case, which could have cosmological implications,
has not yet been established.

\vskip 3mm



\end{abstract}

\thispagestyle{fancy}
\setcounter{page}{0}

\maketitle

\section{Introduction}

Although we are still trying to understand how confinement arises
in QCD, we know that under normal conditions the running coupling
is large at low energies, causing quark and gluons to be confined.
However, when exposed to extreme conditions such as very high temperatures
or densities, quarks are forced to stay at very short distances from
one another and there is a transition to a deconfined, quark-gluon-plasma
phase. This transition is present both at high temperature and at
high density, implying a phase diagram where the hadronic phase exists
only near the origin of the plane defined by the temperature and 
chemical-potential axes.
Of course, one wants to know the location of the transition in order to
study properties of the new phase of matter, but it is also interesting to
determine the nature of this transition. In particular, it is important to 
establish if the transition is a strong one, of first order, involving 
discontinuity in the order parameter, or if it is such that the two phases 
are connected smoothly.
This may have consequences for the understanding of the cosmological QCD 
phase transition, which occurred a few microseconds after the big bang and
formed the hadrons we observe today. In this case the transition lies
closer to the temperature axis and its nature is of direct importance to
determine the types of cosmological relics that can be associated with it.
In particular, a first-order transition would very likely be associated with
the formation of cold dark matter clumps \cite{Schwarz:2003du,Hindmarsh:2005ix}.
The nature of the transition must be also taken into account at relativistic
heavy-ion collision experiments. An additional requirement in this
case is the description of dynamic effects \cite{Hama:2004rr}.

The task of studying the QCD phase transition theoretically must be carried 
out by nonperturbative methods and a natural choice is to consider the 
lattice regularization as a formulation of the theory. 
In fact, lattice-QCD simulations allow a nonperturbative description of
the phase transition in hadronic matter at high temperatures and there has been
some recent progress in the description of the transition also in the case
of finite density \cite{Philipsen:2005mj}.
In the case of the finite-temperature transition, there is a qualitative
difference when dealing with the full-QCD case (i.e.\ considering dynamic
quarks) or with the so-called quenched case, in which the gluonic effects are
taken into account but sea quarks are taken to be infinitely massive.
For the quenched case one studies the deconfining transition itself,
by means of the order parameter given by the Polyakov loop. The transition in
this case is of first order.
In the full-QCD case there is no equivalent order parameter for the deconfinement
transition and one must consider the chiral phase
transition. This transition occurs when the chiral symmetry --- exact
in the limit of zero quark masses and spontaneously broken at low
temperatures --- is restored at high temperature.
The case of two dynamic quarks, i.e.\ considering dynamic effects
of only two degenerate light-quark flavors, corresponding to the up and down 
quarks, is particularly interesting. 
In this case, if the transition is of second order, 
one would expect to observe universal critical
scaling in the class of the $3d$ $O(4)$ continuous-spin model 
\cite{pisarsky,Rajagopal:1992qz}.
Also, in the continuum limit, simulations using different discretizations
for the fermion fields should give the same results.
The fact that the critical behavior should be in the universality class of a
spin model can be precisely checked, since the nonperturbative behavior for
these models can be obtained with Monte Carlo simulations by so-called {\em global}
methods, which avoid the critical slowing-down present in QCD simulations 
\cite{wolff,multigrid}.

The determination of the correct nature of the transition in the two-flavor
case is one of the present challenges of lattice QCD, as pointed out by 
Wilczek in \cite{Wilczek:2002wi}.
This prediction has been investigated numerically by lattice simulations
for over ten years, yet there is still no agreement about the order of
the transition or about its scaling properties \cite{katz,Philipsen:2005mj}.
More precisely, the predicted $O(4)$ scaling has been observed
in the Wilson-fermion case \cite{iwasaki}, but not in the staggered-fermion
case, believed to be the appropriate formulation for studies of the
chiral region. In this case, extensive numerical studies and scaling tests
have been done by the Bielefeld \cite{karsch}, JLQCD \cite{aoki} and MILC
\cite{bernard} groups. It was found that the chiral-susceptibility peaks
scale reasonably well with the predicted exponents, but no agreement is
seen in a comparison with the $O(4)$ scaling function.
At the same time, some recent numerical studies with staggered fermions
claim that the deconfining transition may be of first order 
\cite{cea,delia}.

In \cite{mendes,Mendes:2002pt} a simple method was introduced to
obtain a uniquely defined normalization of the QCD data,
allowing an unambiguous comparison to the (normalized) $O(4)$ scaling 
function. The analysis showed a surprisingly better agreement for the
{\em larger} values of the quark masses. Let us note that in previous 
scaling tests the comparison had been done up to a (non-universal) 
normalization of the data and a match to the scaling function was 
tried by fitting it to the data points with the smallest masses.
One interpretation of this result
is that data at smaller masses (closer to the physical
values) suffer more strongly from systematic errors in the simulations.
In fact, larger quark masses are much easier to simulate, allowing 
greater control over errors and more reliable results. 
Here we present a preliminary study at
a rather large mass value ($m_q = 0.075$ in lattice units), 
using staggered fermions and the MILC code.
We consider the standard action and temporal lattice extent
$N_{\tau}=4$, as in the studies mentioned above.

\section{Scaling tests}
\label{univ}
The behavior of systems around a second-order phase transition 
(or critical point) may show striking similarities for systems 
that would otherwise seem completely different. In fact, it is
possible to divide systems into so-called universality classes,
in such a way that each class will have, e.g., the same critical
exponents around the transition. Typical exponents are
\begin{eqnarray}
M_{h=0,\,t\to 0^-} &\to & |t|^{\beta}
\mbox{,} \\
\chi_{h=0,\,t\to 0} &\to & |t|^{-\gamma}
\mbox{,} \\
M_{t=0,\,h\to 0} &\to & h^{1/\delta}\,,
\end{eqnarray}
where $M$ is the order parameter --- e.g.\ the magnetization
for a spin system --- $\chi$ is the corresponding susceptibility and
\begin{eqnarray}
t &=& (T-T_c)/T_0, \\
h &=& H/H_0
\end{eqnarray}
are the reduced temperature and magnetic field, respectively.

Thus, in principle, one may compare the critical exponents for
different systems to check if they belong to the
same universality class. In practice, however, the critical exponents
may vary little from one class to the other and in order to carry out
the comparison one would need to have a very precise determination
of the exponents, which is not yet feasible in the QCD case.

A more general comparison is obtained through the {\em scaling functions}
for both systems. This comparison allows a more
conclusive test, and can be applied for cases where the critical 
exponents cannot be established with great accuracy.
In this case we may assume the exponents for a given class and 
compare the behavior of the whole critical region for one system
to the known scaling curve for the proposed universality class.
The scaling Ansatz is written for the free energy $F_s$ in
the critical region as
\begin{equation}
F_s(t,h) \;=\;
b^{-d}\,F_s(b^{y_t}\,t, b^{y_h}\,h)\,,
\label{ansatz}
\end{equation}
where $b$ is a rescaling factor, $d$ is the dimension and
$y_t,y_h$ are related to the usual
critical exponents: $\beta$, $\gamma$, $\delta$ mentioned above.
The scaling Ansatz implies that the order parameter
must be described by a universal function
\begin{equation}
M/h^{1/\delta} = 
f_M(t/h^{1/\beta \delta})\;.
\end{equation}
The statement that the function $f_M$ is {\em universal} means that
once the non-universal normalization constants $T_0$
and $H_0$ are determined for a given system in the universality class,
the order parameter $M$ scales according to the
scaling function $f_M$ for all systems in this class.
As said above, the comparison of (normalized) scaling functions 
between two systems is a more general test of universality, especially 
in the case of the QCD phase transition.

A further difficulty in studying the critical behavior at the QCD phase
transition is the impossibility of considering the critical point directly,
since that would correspond to having zero quark mass, or
zero magnetic field $H$ in the language of the spin models above.
In order to check scaling with critical exponents of a given class,
or to determine the normalization constants $T_0$ and $H_0$
for systems where a study at $H=0$ is not
possible, it is important to determine the
{\em pseudo-critical line}, defined by the points where
the susceptibility $\chi$ shows a (finite) peak. This
corresponds to the rounding of the divergence that would be observed
for $H=0$, $T=T_c$. The susceptibility scales as
\begin{equation}
\chi \,=\, \partial M/\partial H \,=\,
(1/H_0)\,h^{1/\delta - 1} 
\,f_{\chi}(t/h^{1/\beta})\;,
\end{equation}
where
$f_{\chi}$ is a universal function related to $f_M$.
At each fixed $h$ the peak in $\chi$ is given by
\begin{eqnarray}
t_{p} &=& z_p\,h^{1/\beta \delta}, \\
M_p &=& h^{1/\delta}\,f_M(z_p), \\
H_0\,\chi_{p} &=& h^{1/\delta - 1} \, f_{\chi}(z_p)\,.
\end{eqnarray}
Thus, the behavior along the pseudo-critical line is determined by the   
universal constants
$z_p$, $f_M(z_p)$, $f_{\chi}(z_p)$.
Critical exponents, the scaling function $f_M$
and the universal constants above are well-known for the 
$3d$ $O(4)$ model \cite{o4,o4o2,o4new}.

Note that one may also consider the comparison for finite-size-scaling
functions, since they are also universal and have
the advantage of being valid for finite values of $L$, the linear size
of the system. Such functions were studied for the $3d$ $O(4)$ model
in \cite{o4o2}.

\section{Comparison of QCD data with the predicted scaling function}

We now turn to the comparison of the two-flavor QCD data in the critical 
region (in the case of small but nonzero quark mass) to the predicted
scaling properties of the $3d$ $O(4)$ spin model.
As mentioned in the Introduction, we consider the chiral phase transition,
since there is no clear order parameter for the deconfinement transition
in the case of full QCD.
The order parameter for the chiral transition is given by the so-called
chiral condensate $<\overline \psi \,\psi>$, where $\psi$ is a combination
of the quark fields entering the QCD Lagrangian \cite{Rajagopal:1992qz}. 
The analogue of the magnetic field 
is the quark mass $m_q$, and (on the lattice) the reduced 
temperature is proportional to 
\begin{equation}
6/g^2 - 6/g_c^2(0)\,,
\end{equation}
where $g$ is the lattice bare coupling and $g_c^2(0)$ is its extrapolated
critical value.
Therefore, referring to the pseudo-critical line described in the
previous section, the chiral susceptibility peaks at
\begin{equation}
t_{p}\sim {m_q}^{1/\beta\delta}\,.
\end{equation}

As mentioned in the Introduction, previous results from lattice-QCD 
simulations in the two-flavor case show good scaling (with the predicted 
exponents) {\em only} along the pseudo-critical line, which is given by the 
peaks of the chiral susceptibility. It should be clear from the discussion
in the above sections that this is not a sufficient test to prove that
the transition is second order, especially if no agreement is seen
when comparing the data to the scaling function.
As described in \cite{mendes,Mendes:2002pt}, we use the
observed scaling along the pseudo-critical line
and the universal quantities 
$z_p$, $f_M(z_p)$ from the $O(4)$ model
to determine the normalization constants
$H_0$, $T_0$ for the QCD data. This allows an unambiguous 
comparison of the data to the scaling function $f_M$. 
More precisely, we note that in previous analyses \cite{bernard}
the normalization constants were tentatively adjusted
by shifting the $O(4)$ curve so as to get a rough agreement
with the data at smaller quark masses,
since these are closer to the chiral limit. 
The problem is that the lighter masses are also more subject to
the presence of systematic errors in the simulations.
In this case the overall
agreement was rather poor, indicating that there were strong systematic 
effets or that the transition is not in the predicted universality class.
Here we fix the constants as described in Section \ref{univ},
following the behavior along the pseudo-critical line. 
In this way no value of the quark mass is priviledged and the comparison 
is unambiguous. 

Our comparison is shown in Fig.\ \ref{scaling} below.
The pseudo-critical line corresponds to a point in this plot and
is marked with an arrow. For clarity we do not show the data
(from the Bielefeld and JLQCD collaborations) obtained directly at
the pseudo-critical point. These are slightly scattered around $z_p$ but
show good scaling within errors.
\begin{figure*}[htbp]
\includegraphics[height=0.5\hsize,angle=-0]{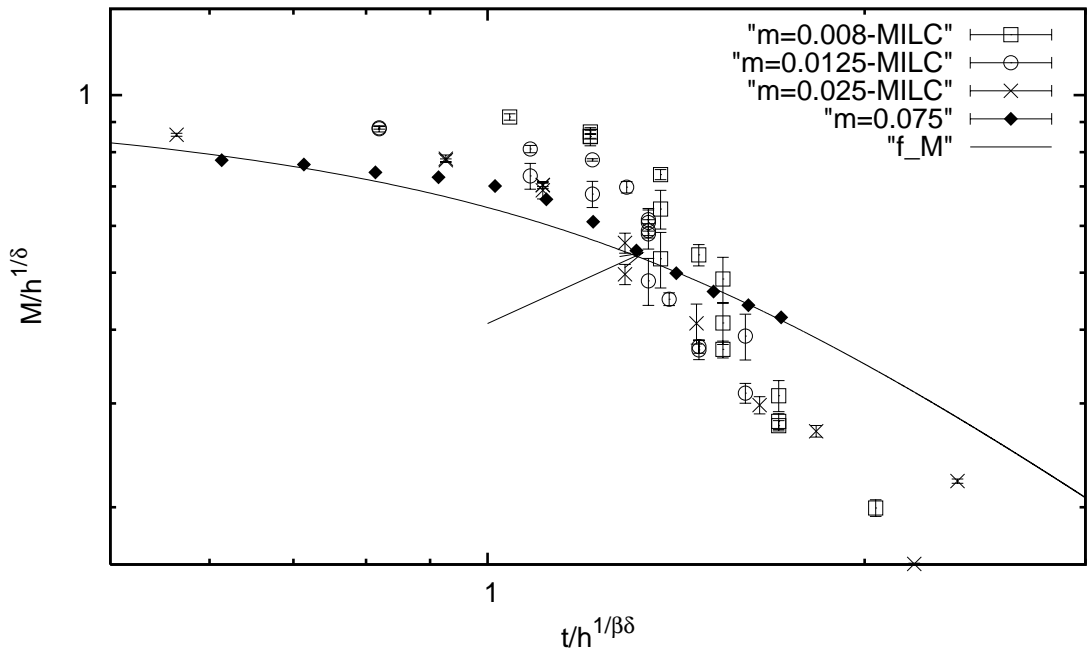}
\caption{
Comparison of QCD (staggered) data to the $O(4)$ scaling
function. For clarity, we do not show the data around
the pseudo-critical point (indicated by the arrow), which were
used to determine the normalization of the remaining data 
points.}
\label{scaling}
\end{figure*}
We see relatively good scaling in the pseudo-critical region,
i.e.\ around [$z_p$, $\,f_M(z_p)$], as expected.
Away from this region most MILC points
are several standard deviations away from the predicted curve.
These data are given for three values of the quark mass in lattice units:
0.008, 0.0125 and 0.025.
Note that the points with larger mass come closer to the curve.
In particular, we can see that the new data at $m_q = 0.075$ show sensibly
better scaling, especially for larger temperatures, where previously
the scaling seemed unlikely.
The good agreement of these data with the $O(4)$ scaling function
motivates a careful study of systematic errors for smaller masses.
A possible source of such errors are finite-size corrections, which would be
stronger for smaller masses, since then the lattice side may not be large
enough to ``contain'' the physical particle.
Put differently, finite-size effects are expected when the correlation length 
(in lattice units) associated with a particle is comparable
to or larger than the lattice side. Of course, this is more likely to occur
for a lighter particle.


\section{Finite-size effects}
\label{FSS}
In addition to the infinite-volume scaling laws mentioned above,
we may also consider finite-size-scaling functions. 
In fact, the scaling Ansatz in Eq.\ \ref{ansatz} also implies
\begin{equation}
M = L^{-\beta/\nu} \,
Q_z(h\,L^{\beta\delta/\nu})
\end{equation}
where $L$ is the linear size of the system
and we consider fixed values of the ratio
$t/h^{1/\beta \delta}
\equiv z$
(e.g.\ $z=0$ as in the critical isotherm,
or $z_p$ as along the pseudo-critical line).
Thus, $M$ can be described by a universal
finite-size-scaling (FSS) function of one variable.
We note that in order to recover the infinite-volume expression
$M=h^{1/\delta}\,f_M(z)$
as $L\to\infty$, we must have
$\;Q_z(u)\,\to\,
f_M(z)\,u^{1/\delta}\;$
for large $u$. Thus, in this limit,
the FSS functions are given simply in terms of
the scaling function $f_M(z)$.
Working with the FSS functions $Q_z$ instead of
the infinite-volume scaling function $f_M$ has the
disadvantage that one must consider $z$ fixed
(thus restricting the regions to be compared in parameter space)
but the advantage that a comparison can be made
already at finite values of $L$.

A finite-size-scaling analysis as described above was carried out in
\cite{mendes}, but it was found that the QCD data show good (finite-size) 
scaling only along the pseudo-critical line.

\section{Conclusions}
Understanding the nature of the chiral phase transition in
two-flavor QCD has proven to be a challenging task. The
prediction of a second-order transition with critical behavior
in the universality class of the $O(4)$ spin model is not verified for
staggered fermions of small masses, although it can be shown (by
an unambiguous normalization of the data) that better scaling is 
obtained for the existing data at larger (unphysical) masses.
The fact that data for heavier quarks would show such good scaling
was not expected, since the normalization of the data for comparison with
the scaling curve did not priviledge any particular values of the quark mass.
This suggests that the lack of scaling at small masses observed so far may be 
due to systematic effects, which could be due to finite-size corrections or to
uncontrolled errors in the hybrid Monte Carlo algorithm used for
updating the configurations. Both these sources of errors would be more
significant for the case of smaller masses.
As discussed above in Section \ref{FSS} above, the deviations observed 
are most likely not due to finite-size corrections and we thus argue that
the deviations from $O(4)$ scaling at smaller
masses may come from systematic errors in the simulation,
probably related to the use of the R algorithm for the simulations 
\cite{clark}. Note that, contrary to what happens for the quenched-QCD case,
the algorithm used to update full-QCD configurations is not exact and
should have its accuracy tested carefully for each different value of the
quark mass used.

Let us also mention that a redefinition of the reduced temperature in terms 
of the physical temperature $T$ including a term in the quark mass $M$,
as suggested in \cite{delia}, improves the agreement with the scaling curve
further, as has been recently shown in \cite{previous2}.

\section{Acknowledgements}
This work was supported by FAPESP (Grant 00/05047-5).
Partial support from CNPq is also acknowledged.


\bibliographystyle{aipproc}   
\bibliography{hadrons_proc}
\IfFileExists{\jobname.bbl}{}
 {\typeout{}
  \typeout{******************************************}
  \typeout{** Please run "bibtex \jobname" to optain}
  \typeout{** the bibliography and then re-run LaTeX}
  \typeout{** twice to fix the references!}
  \typeout{******************************************}
  \typeout{}
 }

\end{document}